\documentclass[twocolumn,pre,showpacs]{revtex4-1}
\usepackage{graphicx}

\begin{document}
\title{Evidence for a bicritical point in the XXZ Heisenberg
  antiferromagnet on a simple cubic lattice}
\author{Walter Selke}
\affiliation{Institut f\"ur Theoretische Physik, RWTH Aachen
  University, 52056 Aachen, Germany}

\begin{abstract}
The classical Heisenberg antiferromagnet with uniaxial exchange
anisotropy, the XXZ model, in a magnetic field on a simple cubic lattice is
studied with the help of extensive
Monte Carlo simulations. Analyzing, especially, various
staggered susceptibilities and Binder cumulants, we present
clear evidence for the meeting point of the antiferromagnetic, 
spin--flop, and paramagnetic phases being a bicritical
point with Heisenberg symmetry. Results are compared to
previous predictions based on various theoretical 
approaches.
\end{abstract}

\pacs{75.10.Hk, 75.40.Cx, 05.10.Ln}

\maketitle
Uniaxially anisotropic Heisenberg antiferromagnets in a
magnetic field have been
studied quite extensively in the past, both experimentally
and theoretically \cite{rev1,rev2}. Usually, they display, at low
temperatures and fields, the antiferromagnetic phase and, when increasing
the field, the spin--flop phase. A prototypical model
describing these phases as well as, possibly, multicritical
points, is the Heisenberg model with a
uniaxial exchange anisotropy, the XXZ model
 \begin{equation}
  {\cal H} = J \sum\limits_{i,j}
  \left[ \, \Delta (S_i^x S_j^x + S_i^y S_j^y) + S_i^z S_j^z \, \right]
  \; - \; H \sum\limits_{i} S_i^z
\end{equation}
where $J(> 0)$ is the exchange coupling between classical
spins, $(S_{i(j)}^x,S_{i(j)}^y,S_{i(j)}^z)$, of length one
at neighboring sites, $i$ and
$j$, on a simple cubic lattice, $\Delta$ is the uniaxial exchange
anisotropy, $1 > \Delta > 0$, and $H$ is the applied
magnetic field along the easy axis,
the $z$--axis. 
The phase diagram of the model has been investigated already several years
ago, using, among others, mean--field theory \cite{Gorter}, Monte
Carlo (MC) simulations \cite{LanBin}, and high temperature series
expansions \cite{Mourit}. The transition between the
antiferromagnetic (AF) and spin--flop (SF) phases seems to be of
first order, while the boundaries of the paramagnetic (P) phase to the AF
and SF phases are believed to be continuous transitions in the Ising
and XY universality classes, respectively. Moreover, based on
renormalization group analyses
in one--loop--order, a bicritical point in the Heisenberg 
universality class had been proposed, at which the
three different phases meet \cite{FN,KNF}. This
scenario has been questioned on the basis of
renormalization group calculations in high--loop--order \cite{CPV}, where the
bicritical point has been argued to be unstable against
a tetracritical point \cite{Aha}, which, in turn, may be unstable
towards transitions of first order in the vicinity of
the meeting point of the three phases. However, a subsequent renormalization
group analysis in two--loop--order
\cite{Folk} suggests that a bicritical point in
the Heisenberg universality class can not be excluded.

\begin{figure}
\resizebox{0.85\columnwidth}{!}{%
  \includegraphics{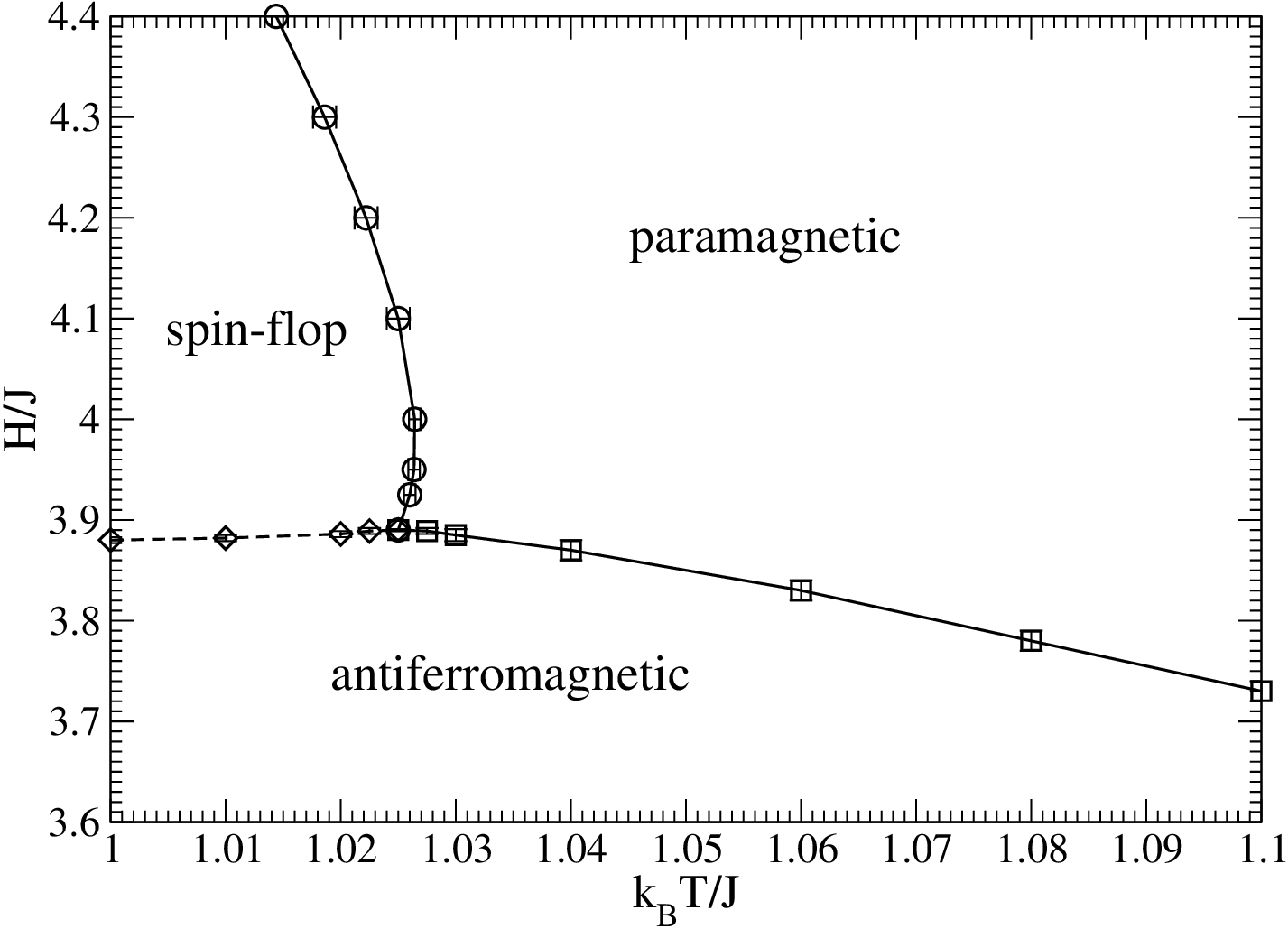}
}
  \caption{Phase diagram of the XXZ model on a cubic lattice
    with exchange anisotropy $\Delta= 0.8$ in the vicinity
    of the AF--SF--P point, as obtained from
    the present MC simulations.}
\label{fig:1}
\end{figure}

As has been noted quite recently \cite{HWS}, not only
AF and SF phases, but also biconical (BC) structures \cite{KNF} 
may play an important role in the XXZ model. Indeed, such
BC structures are degenerate ground states at the critical field 
separating AF and SF configurations at zero temperature. For the
XXZ model on a {\it square} lattice, these degenerate BC fluctuations
lead to a narrow disordered phase intervening between
the AF and SF phases at low temperatures, giving, presumably, rise to
a 'hidden tetracritical
point' \cite{HWS,Zhou} at zero temperature. A recent Monte
Carlo study \cite{bann} for the XXZ antiferromagnet on
a {\it simple cubic} lattice showed that biconical structures, arising
from the degenerate ground states, also occur at low temperatures
close to the transition between the AF and SF phases. But they
do not destroy the direct transition of first order between
these two phases, in accordance with the behavior predicted
by mean--field and other, more reliable theories. 

In that recent MC study \cite{bann} of the three--dimensional XXZ model,
Eq. (1), at fixed anisotropy, $\Delta =0.8$, continuous phase
boundaries of Ising type, for the AF--P transition, and of XY
type, SF--P, as well as the transition line of first order, AF--SF, have
been identified. The position of the meeting (or AF--SF--P) point has
been estimated, $k_BT/J= 1.025 \pm 0.015$ and
 $H/J= 3.9 \pm 0.03$, without determining critical properties of the AF--SF--P
point. The aim of the present paper is to deal with this intriguing
aspect, using extensive MC simulations.

Here, the standard Metropolis algorithm \cite{LanBinMC} with single
spin--flips is applied. Employing full periodic boundary
conditions, lattices with $L^3$ sites, are considered. $L$ ranges from
4 to 40, with the main focus on the sizes $L$= 8, 16, 24, and 32, to
study systematically finite--size effects. MC runs with, at least, $10^7$
MC steps per site for the larger lattices, are performed. Error
bars are estimated by averaging over, at least, three independent
realizations. In this way, data of the desired accuracy are obtained,
and there is no need to use other, perhaps, more powerful 
MC algorithms.

To map the phase diagram, we study, especially, quantities related
to the Ising, XY, and Heisenberg order parameters. In
particular, we monitor, apart from the absolute values 
of the staggered longitudinal, $m^z_{st}$, transverse,
$m^{xy}_{st}$, and isotropic, $m^{xyz}_{st}$, magnetizations, the 
corresponding staggered susceptibilities, 
$\chi ^z_{st}, \chi ^{xy}_{st}$, and $\chi^{xyz}_{st}$, as well as the
 corresponding Binder cumulants
\cite{Binder}, $U^z, U^{xy}$, and $U^{xyz}$. To our knowledge, the
isotropic quantities, being crucial to identify a bicritical point
with Heisenberg symmetry, had not been included in previous MC
simulations.

\begin{figure}
\resizebox{0.88\columnwidth}{!}{%
  \includegraphics{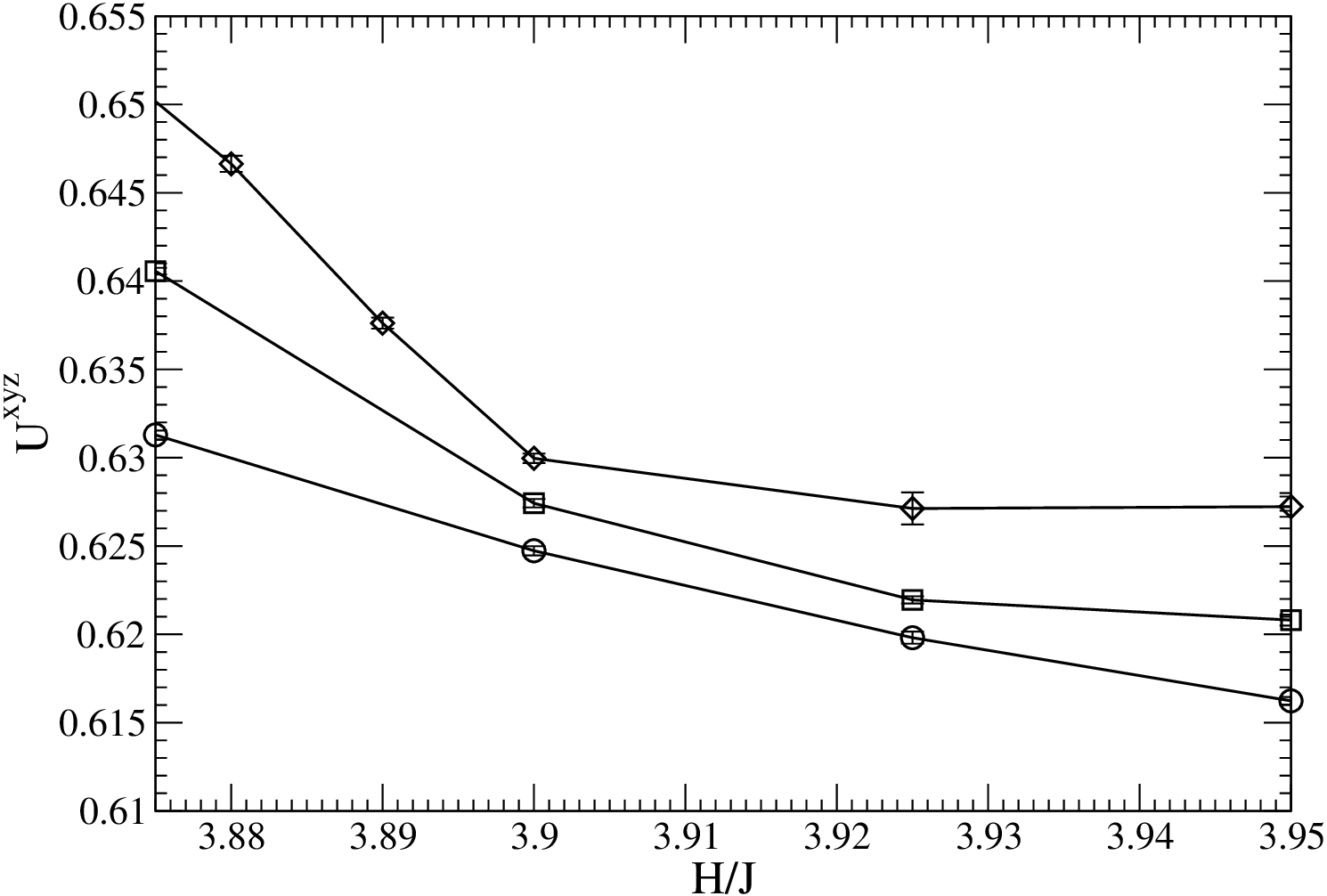}
}
\caption{Isotropic Binder cumulant $U^{xyz}$, at $k_BT/J= 1.0225$, varying
the field, for lattice sizes $L= 16$ (circles), 24 (squares), and
32 (diamonds). }
\label{fig:2}
\end{figure}
\vspace{1.5cm}
\begin{figure}
\resizebox{0.88\columnwidth}{!}{%
  \includegraphics{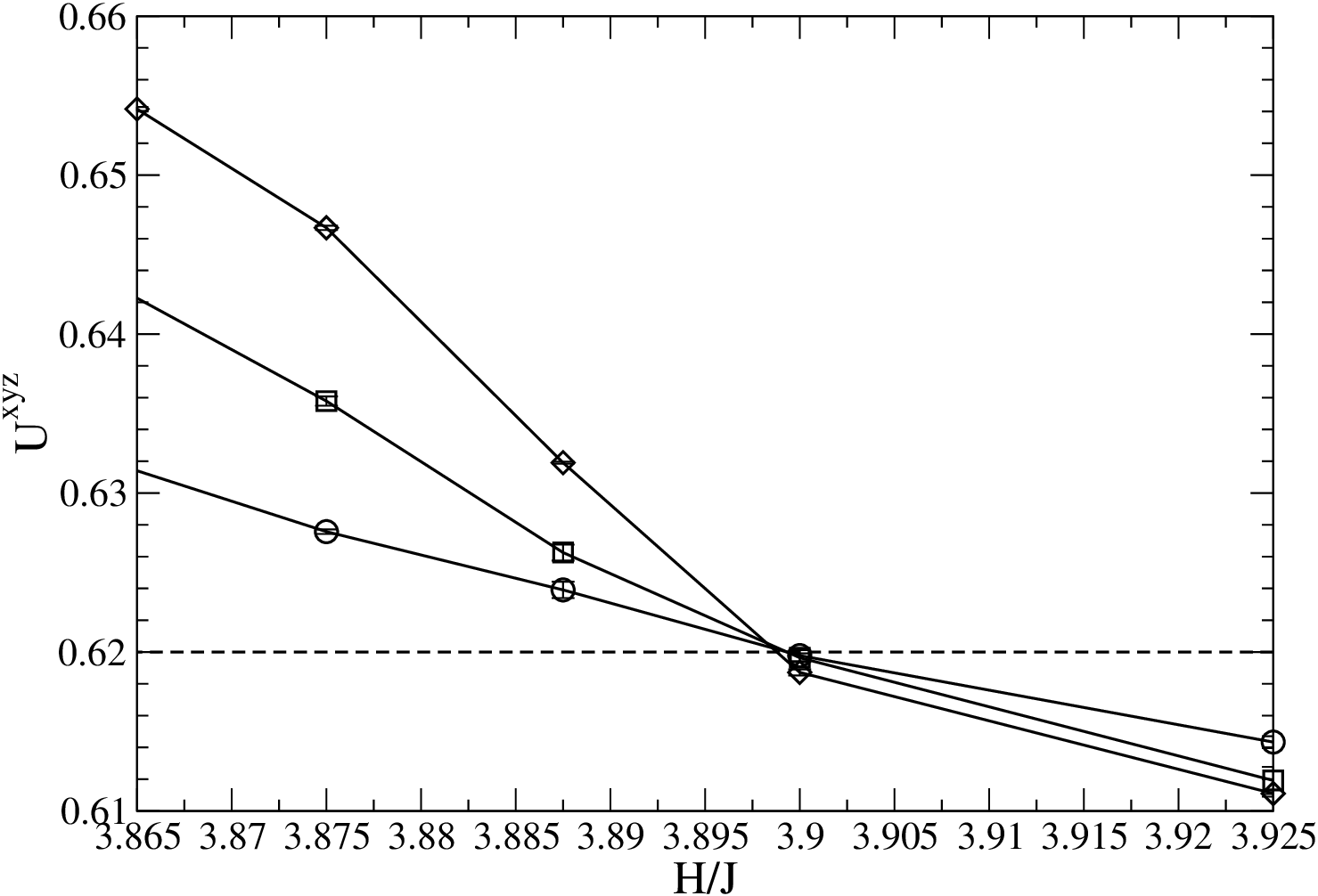}
}
  \caption{As for Fig. 2, at $k_BT/J= 1.025$, including $(U^{xyz})^*$
    as dashed line \cite{hase2,pec}.}
\label{fig:3}
\vspace{0.8cm}
\end{figure}

\vspace{-1.4cm}

The phase diagram, in the ($k_BT/J, H/J$)--plane, in the
vicinity of the AF--SF--P point is depicted in Fig.1. As before, we
set $\Delta$= 0.8. The
shown estimates for the transition lines are based on
usual finite--size analyses \cite{Barber}. In particular, to
determine the AF--P boundary, with the transition being in the
Ising universality class, the size dependence of the height
and position of the maximum in the longitudinal staggered
susceptibility, $\chi^z_{st}$, allows for reliable
estimates. $U^{xy}$ turns out to be very useful in estimating the SF--P
transition line. There one observes 
rather small finite--size correction terms to the critical Binder
cumulant for cubic lattices in the XY universality
class \cite{hase1}, $(U^{xy})^*$= 0.586.., with the critical
Binder cumulant being the cumulant at the transition in
the thermodynamic limit, $L \rightarrow \infty$. Finally, the
AF--SF phase boundary of first order is readily identified from the
location of the maxima in $\chi^z_{st}$ and $\chi^{xy}_{st}$. 

The resulting phase diagram confirms and refines 
previous, independent MC findings \cite{bann}. Note that the present
results allow to locate the AF--SF--P point
accurately, $k_BT/J= 1.025 \pm 0.0025$ and $H/J= 3.89 \pm 0.01$,
reducing appreciably the error bars obtained in the previous
MC study, as mentioned above. 

\begin{figure}
\resizebox{0.88\columnwidth}{!}{%
  \includegraphics{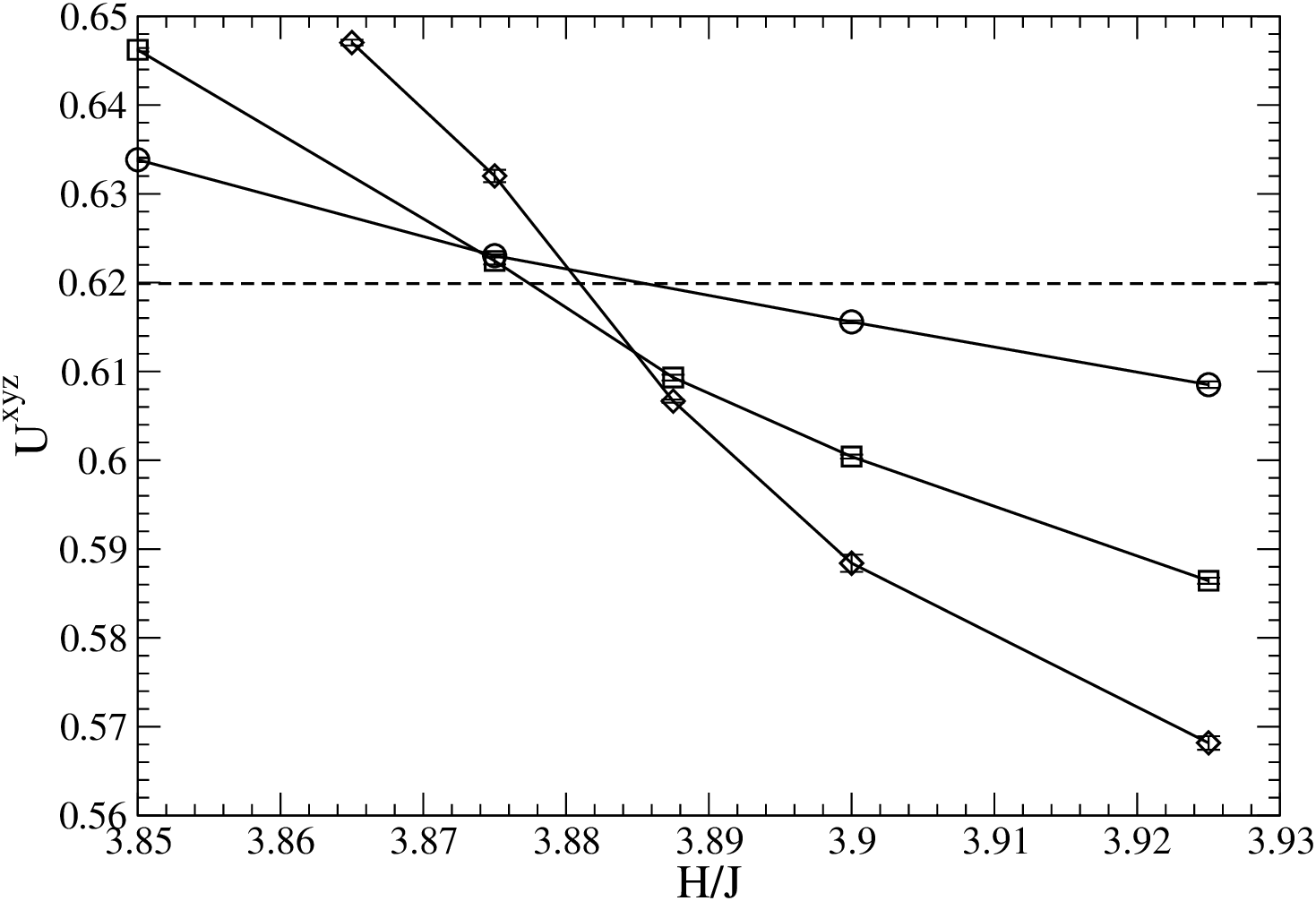}
}
  \caption{As for Fig. 3, at $k_BT/J= 1.03$.}
\label{fig:4}
\end{figure}
Most importantly, critical properties of the AF--SF--P point 
are studied. In case of a bicritical point with Heisenberg symmetry,
the critical Binder cumulant $(U^{xyz})^*$ is expected
\cite{hase2,pec} to acquire the value 0.620 ($\pm 0.003$), using
periodic boundary conditions for lattices of cubic shape. Note
that, in general, the critical Binder cumulant, in
a given universality class, may depend on
boundary condition, system shape, and lattice anisotropy of
the interactions \cite{Binder,CD,SeSh}.

\begin{figure}
\resizebox{0.88\columnwidth}{!}{%
  \includegraphics{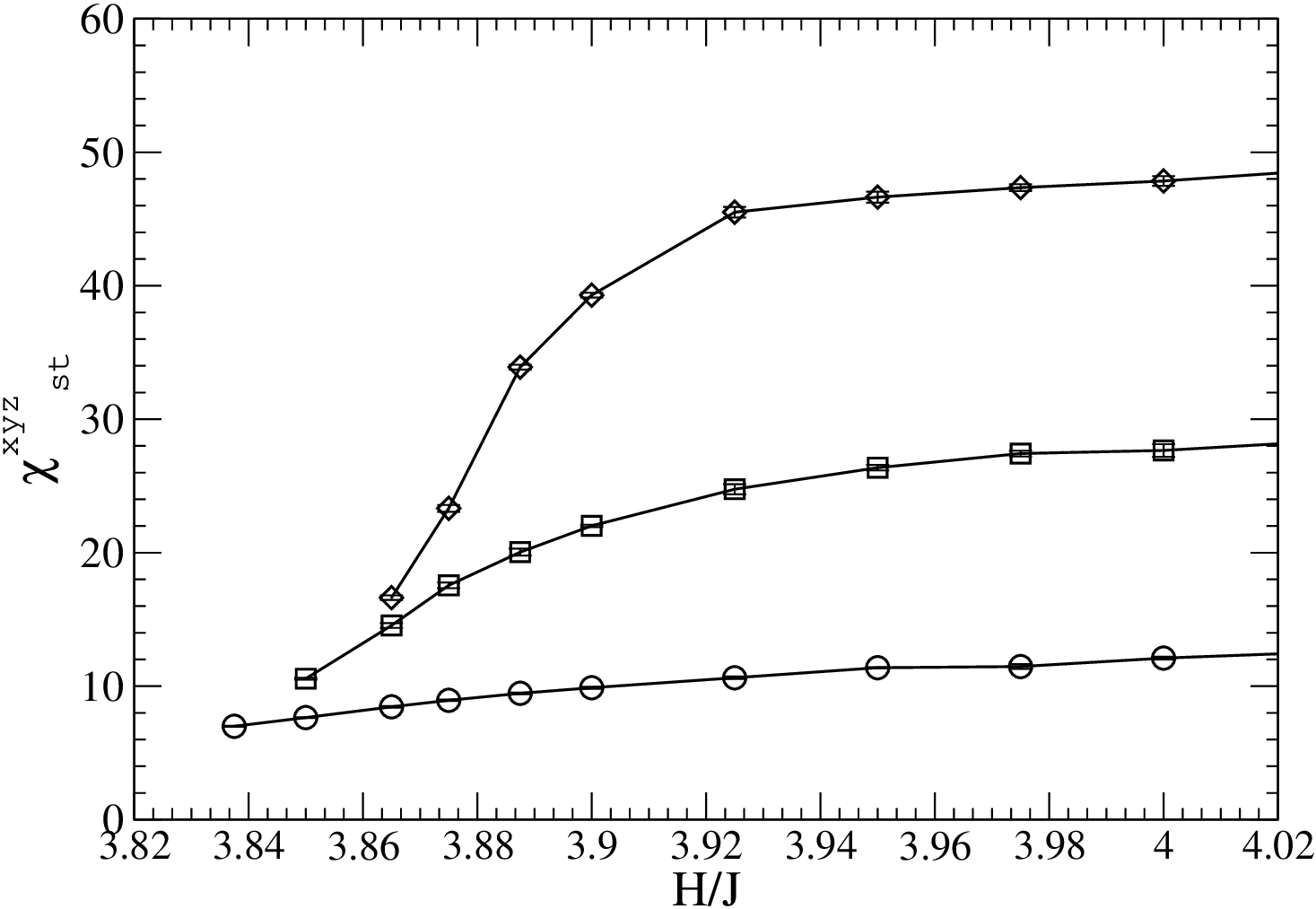}
}
\caption{Staggered isotropic susceptibility $\chi^{xyz}_{st}$ at
  $k_BT/J= 1.025$ for $L$= 16 (circles), 24 (squares), and
  32 (diamonds).}
\label{fig:5}
\vspace{0.8cm}
\end{figure}
\begin{figure}
\resizebox{0.88\columnwidth}{!}{%
  \includegraphics{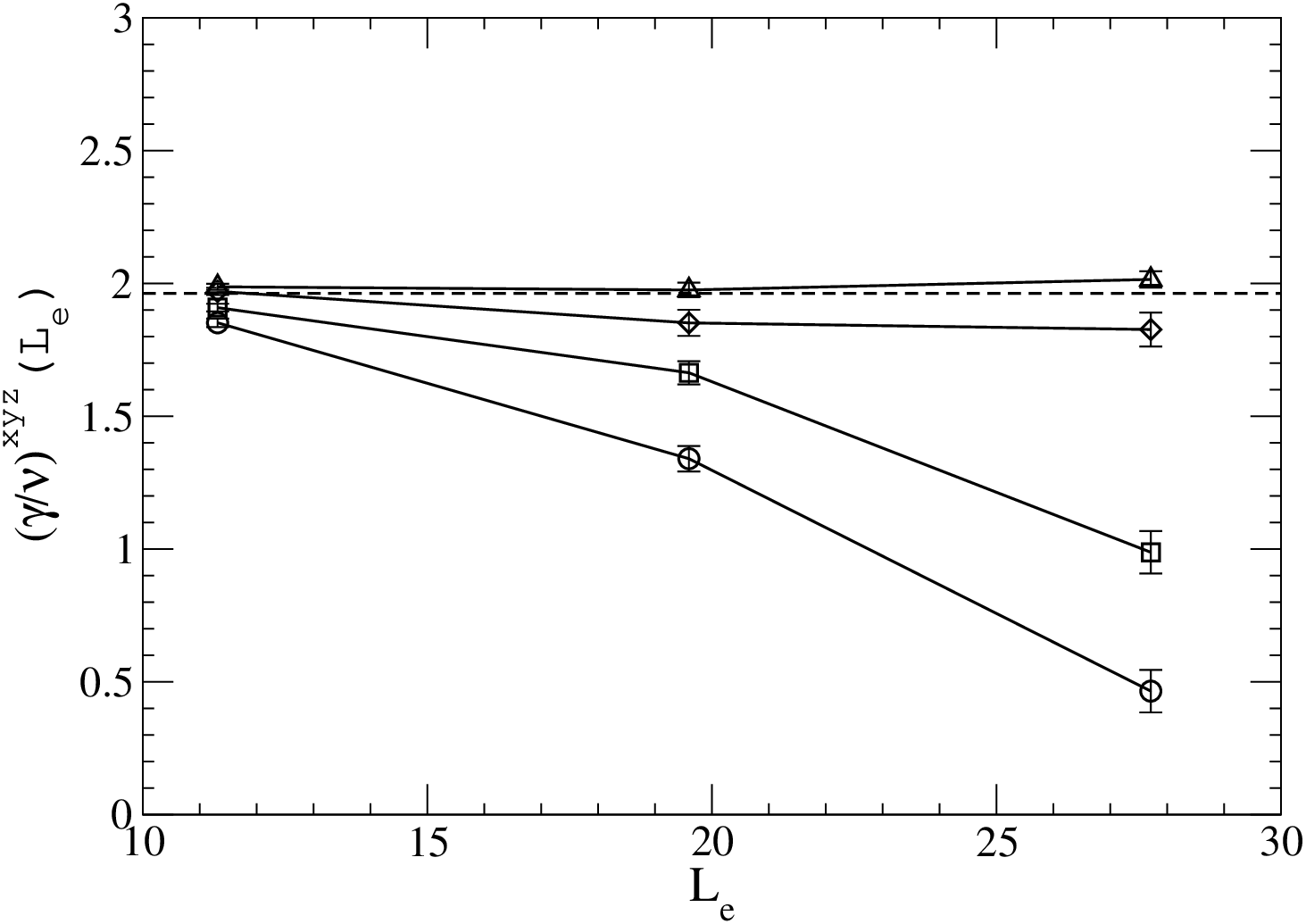}
}
\caption{Effective critical exponent
    $(\gamma/\nu)^{xyz}(L_e)$ 
    of the staggered isotropic susceptibility, see Eq. (2), at
    $k_BT/J= 1.025$ and
    $H/J$= 3.865 (circles), 3.875 (squares), 3.8875 (diamonds), and
    3.90 (triangles). The dashed line marks the  asymptotic critical
    exponent in the Heisenberg universality class \cite{PV}. }
\label{fig:6}
\end{figure}

In Figs. 2, 3, and 4 the size dependent isotropic Binder
cumulant $U^{xyz}$ is displayed near the AF--SF--P
point, at three fixed temperatures, $k_BT/J$= 1.0225, 1.025, and
1.03, varying the field to cross the boundary of the AF phase, see
also Fig. 1. We may distinguish three different scenarios. At 
the lowest temperature, $k_BT/J= 1.0225$, see Fig. 2, the cumulant increases 
with larger system sizes at all fields, so that there are
no intersection points for cumulants of different
sizes. This behavior is in accordance with
being at a temperature below that of the AF--SF--P
point. Indeed, $U^{xyz}$ tends to 2/3 in the AF and SF phases in
the thermodynamic limit, and there is no indication of a transition of
Heisenberg symmetry. At $k_BT/J= 1.025$, see Fig. 3, all
intersection points of the cumulants, for lattices of
sizes $L$= 16, 24, and 32, occur closely to the
critical Heisenberg value,  $(U^{xyz})^* \approx 0.620$. This
fact may be interpreted as evidence for being in the immediate 
vicinity of a bicritical point of Heisenberg symmetry. Actually,
moving to higher temperatures, another scenario shows up. There
are still intersection points of the cumulants for different
lattice sizes, but they shift for larger lattices
to lower and lower values below that of the critical cumulant in
the Heisenberg universality class. This trend is already
seen for $k_BT/J= 1.0275$, and it is quite pronounced at
$k_BT/J= 1.03$, as depicted in Fig. 4. In any event, the
observations on the isotropic Binder cumulant are completely
consistent with the existence of a bicritical point with
Heisenberg symmetry at $k_BT/J= 1.025 \pm 0.0025$. 

This suggestion is supported by the behavior of the isotropic
staggered susceptibility $\chi^{xyz}_{st}$. Simulation data
at fixed $k_BT/J= 1.025$ are shown in Fig. 5. At first sight, there seems to
be no hint for criticality of Heisenberg type, because
$\chi^{xyz}_{st}$ as a function of field exhibits no
maximum. However, additional information is provided by
the standard discrete effective critical exponent, at given
temperature and field, 

 \begin{equation}
  (\gamma/\nu)^{xyz}(L_e)= \ln(\chi^{xyz}_{st}(L_1)/\chi^{xyz}_{st}(L_2))/\ln(L_1/L_2)
\end{equation}

\noindent
for successive sizes $L_1>L_2$ and with the
effective length $L_e= \sqrt{L_1L_2}$. In case of
critical behavior in
the Heisenberg universality class, the effective exponent would
approach, for $L_e \longrightarrow \infty$, the
known \cite{PV} asymptotic value $(\gamma/\nu)^{xyz}$= 1.96... In
fact, as shown in Fig. 6, the effective exponent seems to approach 
closely that value at $H/J= 3.8875 \pm 0.0125$. Obviously, 
at those fields one is in a (crossover) critical
region governed by the Heisenberg
fixed point, in accordance with the findings for the isotropic
Binder cumulant. The suggestion is corroborated by the behavior of
the staggered longitudinal and transverse susceptibilities. Both
quantities show pronounced maxima, when varying the field
at $k_BT/J$= 1.025. The size dependence of the peak position
allows us to estimate the critical field. Within the error bars, both
susceptibilities lead to the same critical
field, $H/J= 3.895 \pm 0.005$, indicating, again, the closeness
of the bicritical point.

It should be mentioned that we did, in addition, a preliminary MC
study at a rather strong anisotropy, $\Delta= 0.2$. The AF--SF--P point
is estimated to be at about $k_BT/J= 0.2095 \pm 0.0005$ and
$H/J= 5.825 \pm 0.0025$. In the immediate vicinity of that point, the
isotropic Binder cumulant shows similar features to the ones
discussed above. We tend to believe that the AF--SF--P point
in the XXZ Heisenberg 
antiferromagnet on a simple cubic lattice is generically a
bicritical point with Heisenberg symmetry.

I should like to thank R. Folk, G. Bannasch, T.--C. Dinh, and
D. Peters for very useful discussions.


\begin{thebibliography}{}

\bibitem{rev1}  Y.\ Shapira, in {\it Multicritical Phenomena} , ed. by
  R.\ Pynn and A. Skjeltorp (Plenum Press, New York and London, 1984),
  p.35; and references therein.
\bibitem{rev2} W.\ Selke, M.\ Holtschneider, R.\ Leidl, S.\ Wessel,
  G.\ Bannasch, and D.\ Peters, Physics Procedia \textbf{6}, 84
  (2010); and references therein.
\bibitem{Gorter} C.\ J.\ Gorter and T.\ Van Peski-Tinbergen,
  Physica \textbf{22}, 273 (1956).
\bibitem{LanBin} D.\ P.\ Landau and K.\ Binder, Phys.\ Rev.\ B\
  \textbf{17}, 2328 (1978).
\bibitem{Mourit} O.\ G.\ Mouritsen, E.\ K.\ Hansen, and S.\ J.\ K.\
  Jensen, Phys.\ Rev.\ B\
  \textbf{22}, 3256 (1980).
\bibitem{FN} M.\ E.\ Fisher and D.\ R.\ Nelson, Phys. Rev. Lett. {\textbf{32}},
    1350 (1974).
\bibitem{KNF} J.\ M.\ Kosterlitz, D.\ R.\ Nelson, and  M.\ E.\ Fisher, Phys. Rev. B \textbf{13}, 412 (1976).
\bibitem{CPV} P.\ Calabrese, A.\ Pelissetto, and E.\ Vicari,
  Phys.\ Rev.\ B\ \textbf{67}, 054505 (2003).
\bibitem{Aha} A.\ Aharony,
  J.\ Stat.\ Phys.\ \textbf{110}, 659 (2003).
\bibitem{Folk} R.\ Folk, Yu.\ Holovatch, and G.\ Moser,
  Phys.\ Rev.\ E \textbf{78}, 041124 (2008).
\bibitem{HWS} M.\ Holtschneider, S.\ Wessel, and W.\ Selke,
  Phys.\ Rev.\ B\ \textbf{75}, 224417 (2007).
\bibitem{Zhou} C.\ G.\ Zhou, D.\ P.\ Landau, and T.\ C.\ Schulthess,
  Phys.\ Rev.\ B\ \textbf{76}, 024433 (2007).
\bibitem{bann} G.\ Bannasch and W.\ Selke,
  Eur.\ Phys.\ J.\ B\ \textbf{69}, 439 (2009).
\bibitem{LanBinMC} D.\ P.\ Landau and K.\ Binder, {\it A Guide to
    Monte Carlo  Simulations in Statistical Physics} (University
  Press, Cambridge, 2005).
\bibitem{Binder} K.\ Binder, Z. Physik B- Cond. Matt. \textbf{43}, 119 (1981).
\bibitem{Barber} M.\ N.\ Barber, in {\it Phase Transitions and
    Critical Phenomena} , ed. by C.\ Domb and J.\ L.\ Lebowitz (Academic
  Press, New York, 1983), Vol. 8.
\bibitem{hase1} M.\ Hasenbusch and T.\ T\"or\"ok, J. Phys. A-
  Math. Gen. \textbf{32}, 6361 (1999).
\bibitem{pec} P.\ Peczak, A.\ M.\ Ferrenberg, and D.\ P.\ Landau,
  Phys.\ Rev.\ B\ \textbf{43}, 6087 (1991).
\bibitem{hase2} M.\ Hasenbusch, J. Phys. A- Math. Gen. \textbf{34}, 8221 (2001).
\bibitem{CD} X.\ S.\ Chen and V.\ Dohm, Phys.\ Rev.\ E \textbf {70}, 056136
  (2004).
\bibitem{SeSh} W.\ Selke and L.\ N.\ Shchur, J.\ Phys.\ A- Math. Gen. \textbf{38}, L739 (2005).
\bibitem{PV} A.\ Pelissetto and E.\ Vicari, Phys.\ Rep.\ \textbf{368},
  549 (2002).
\end{thebibliography}
\end{document}